\pdfoutput=1

\documentclass[two-column,jgrga]{agutex}

\usepackage{color}
\usepackage{rotating}

 \usepackage{lineno}

\usepackage{graphicx}

\usepackage{amsmath, amssymb}
\authorrunninghead{{\O}STVAND, RYPDAL AND RYPDAL}

\titlerunninghead{SIGNIFICANCE OF TEMPERATURE TRENDS}

\begin{document}

\title{Statistical significance of rising and oscillatory trends  in global ocean and land temperature in the past 160 years}
\authors{L. {\O}stvand \altaffilmark{1}, K. Rypdal \altaffilmark{2}, and M. Rypdal \altaffilmark{2} }

\altaffiltext{1}{Department of Physics and Technology, University of Troms{\o}, Norway.}
\altaffiltext{2}{Department of Mathematics and Statistics, University of Troms{\o}, Norway.}
\authoraddr{L. {\O}stvand, 
(lene.ostvand@uit.no)}

\begin{abstract}
Various  interpretations of the notion of a trend in the context of global warming are discussed, contrasting the difference between viewing a trend as the deterministic response to an external forcing and viewing it as a slow variation which can be separated from the background spectral continuum of long-range persistent climate noise. The emphasis in this paper is on the latter notion, and a general scheme is presented  for testing a multi-parameter trend model against a null hypothesis which models the observed climate record as an autocorrelated noise. The scheme is employed to the instrumental global sea-surface temperature record and the global land-temperature record. A trend model comprising a linear plus an oscillatory trend with period of approximately 60 yr, and the statistical significance of the trends, are tested against three different null models:  first-order autoregressive process, fractional Gaussian noise, and fractional Brownian motion. The linear trend is significant in all cases, but the oscillatory trend is insignificant for ocean data and barely significant for land data. By means of a Bayesian iteration, however, using the significance of the linear trend to formulate a sharper null hypothesis, the oscillatory trend in the land record appears to be statistically significant. The results suggest that  the global land record may be  better suited for detection of the global warming signal than the ocean record.
\end{abstract}

\begin{article}

\section{Introduction}
At the surface of things, the conceptually simplest approach to detection of  anthropogenic global warming should be the estimation of trends in global surface temperature throughout the instrumental observation era starting in the mid-nineteenth century. These kinds of estimates, however,  are subject to deep controversy and confusion originating from disagreement about how the notion of a trend should be understood.  In this paper we adopt the view that there are several, equally valid, trend definitions. Which one that will prove most useful depends on the purpose  of the analysis and the availability and quality of observation data.

At the core of the global change debate is how to distinguish anthropogenically forced warming from natural variability. A complicating factor here is that natural variability has  forced as well as  internal components. Power spectra of climatic time series also suggest to separate internal dynamics into quasi-coherent oscillatory modes and a continuous and essentially scale-free spectral background. Over  a vast range of  time scales this background takes the form of a persistent, fractional noise or motion \citep{Lovejoybook,Markonis2013}.  Hence, the issue is threefold: (i)  to distinguish the climate response to anthropogenic forcing  from the response to natural forcing,   (ii) to distinguish internal dynamics from forced responses, and   (iii) to distinguish quasi-coherent, oscillatory modes from the persistent-noise background. This conceptual structure is illustrated by the Venn diagram in Figure 1(a). Figure 1(b) illustrates three possible trend notions based on this picture. Fundamental for all is the separation of the observed climate record into a trend component (also termed the {\em signal}) and a {\em climate noise} component. The essential difference between these notions is  how to make this separation. 

\begin{figure}
\noindent \includegraphics[width=.49\textwidth]{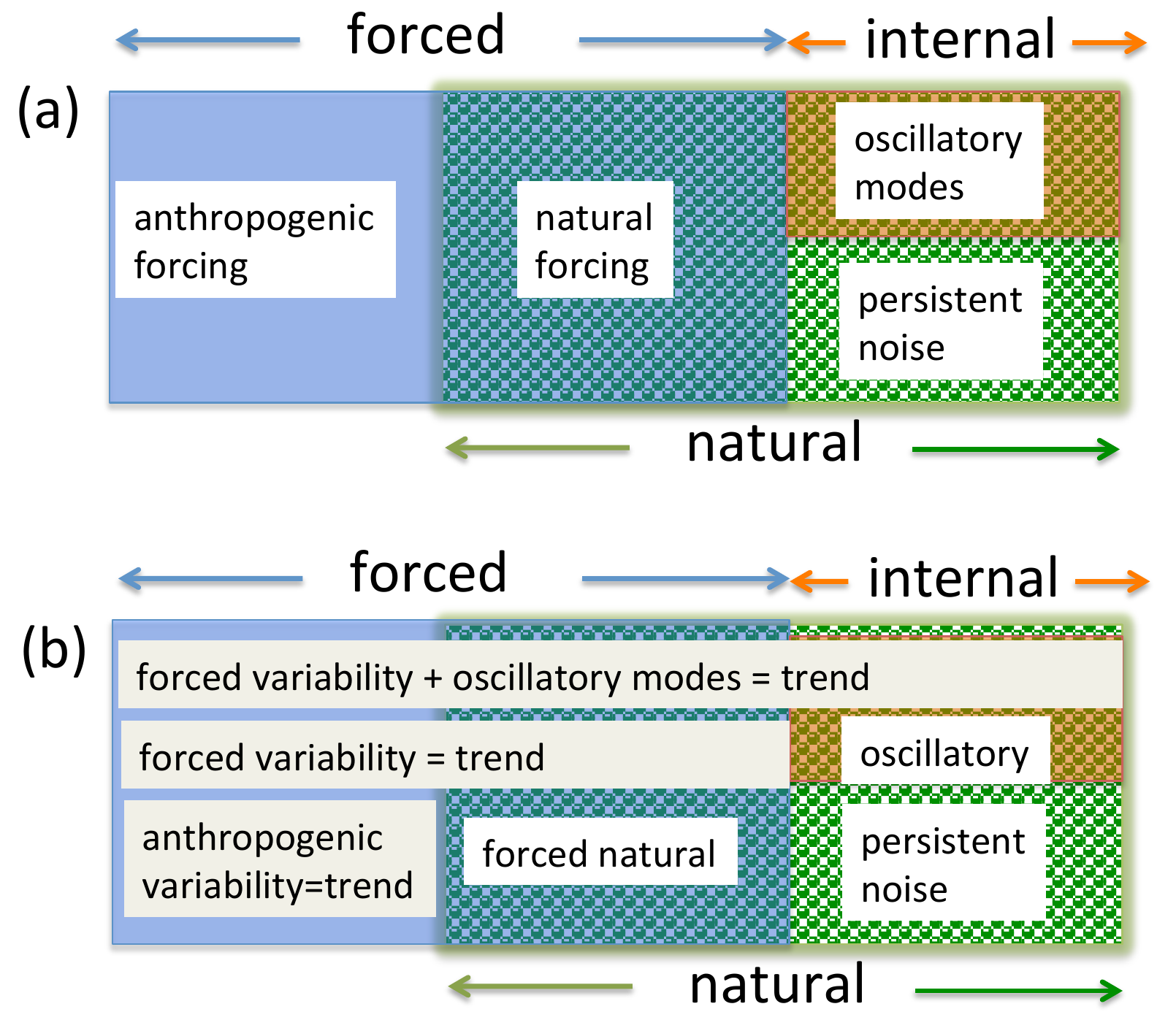}
\caption{Venn diagrams illustrating the interplay between forced, internal, and natural variability and various definitions of trend. (a): Natural variability can be both forced and internal. Forced variability can be both anthropogenic and natural. Internal variability is natural, but can consist of quasiperiodic oscillatory modes as well as a continuum of persistent noise. (b): The three different trend notions discussed in the text.}
\label{fig:Venndiagram}
\end{figure}

The widest definition of the trend is to associate it with all forced variability and oscillatory modes as illustrated by the upper row in Figure 1(b). With this notion the methodological challenge will  be to develop a systematic approach to extract the persistent noise component from the observed record, and then to  subtract this component to establish the trend. The physical relevance of this separation will depend on to what extent we can justify to interpret the extracted trend as a forced response with internally generated oscillatory modes superposed. If detailed information on the time evolution of the climate forcing is not used or is unavailable such a justification is quite difficult. In this case we will first construct a parametrized model for the trend based on the appearance of the climate record at hand and our physical insight about the forcing and the  nature of the dynamics. The next step will be to estimate the parameters of the trend model by conventional regression analysis utilizing the observed climate record. The justification of interpreting this trend as something forced and/or coherent different from background noise will be done through a test of the null hypothesis  which states that the  climate record can be modeled  as a long-range memory (LRM) stochastic process.  Examples of such processes are  persistent fractional Gaussian noises (fGns) or  fractional Brownian motions (fBms). For comparison we will also test the null hypothesis against a conventional short-memory notion of climate noise, the first-order autoregressive process (AR(1)). Rejection of this null hypothesis will be taken  as an acceptance of the hypothesis that the estimated trend is significant, and will strengthen our confidence that these trends represent identifiable dynamical features of the climate system.

It could be argued that the value of  this kind of analysis of statistical significance is of little interest since the result depends on the choice of null model for the climate noise. One can, however, test the  null models against the observation data, and here our analysis seems to favor the fGn/fBm models over short-memory models.

If forcing data  are available over the time span of the observed temperature record we can utilize this information in a parametrized, linear, dynamic-stochastic model for the climate response. The trend then corresponds to the deterministic solution to this model, i.e., the solution with the known (deterministic) component of the forcing. In this model the persistent-noise component of the temperature record is the response to a white-noise stochastic forcing. The method is described in \cite{RR2013}, where only exponential and  scale-free long-range persistent responses are modeled, without allowing for quasi-coherent oscillations. The approach in that paper adopts the trend definition described in the second row of Figure 1(b). Here the trend is the forced variability, while all unforced variability is relegated to the realm of climate noise. It is possible, however, to incorporate forced and natural oscillatory dynamics into such a response model. The simplest way could be to add the response of a forced, damped harmonic oscillator to the scale-free response. These extra degrees of freedom would add an oscillatory  response to the deterministic forcing (this would be a forced, oscillatory response), but also an oscillatory response to the stochastic forcing which would be interpreted as an internal oscillatory mode.  
 According to the approach described in \citep{RR2013} we have to classify all  deterministic forced responses as trends, implying that a trend defined this way is not necessarily slow. For instance, the irregular  sequence of volcanic eruptions provides a shot-noise like forcing signal. After having estimated the parameters of the forced response model using the full forcing data and the observed temperature record, the residual can be analyzed to assess the validity of different noise models. The responses to fast components in the forcing (like volcanic spikes) will be shifted to the forced response, rather than being incorrectly represented as parts of the internal noise. The test of different noise models via analysis of the residual will therefore give more correct results in the forced-response model  than the trend-fit approach employed in the present paper.

The lower row in Figure 1(b) depicts the trend notion of foremost societal relevance; the forced response to anthropogenic forcing. Once we have estimated the parameters of the forced-response model, we can also compute the  deterministic response to the anthropogenic forcing separately. One of the greatest advantages of the forced-response methodology  is that it allows estimation of this anthropogenic trend/response and prediction of future trends under given forcing scenarios, subject to rigorous estimates of uncertainty. This will be a topic for a forthcoming paper.

\section{Trend Detection Without Forcing Data}\label{seclot}
The noise modeling in this paper makes use of the concept of  long-range memory (LRM), or (equivalently) long-term persistence (LTP) \citep{beran1994}.  In global temperature records this has been studied in e.g., \citet{pelletier1999,lennartz2009,rybski2006,Rypdal2010,efstathiou2011,rypdalJGR2013, RR2013}. Emanating from these studies is the recognition that ocean temperature is more persistent than land temperature and that the 20'th century rising trend is stronger for land than for ocean. LRM is characterized by a time-asymptotic ($t\rightarrow \infty)$ autocorrelation function (ACF) of power-law form $C(t)\sim t^{\beta-1}$ for which the integral $\int_0^\infty C(t) \mathrm{d}t$ diverges. Here $\beta$ is a power-law exponent indicating the degree of persistence. The corresponding asymptotic ($f\rightarrow 0$) power spectral density (PSD) has the form $S(f)\sim f^{-\beta}$, hence $\beta$ is also called the spectral index of the LRM process. For $0<\beta<1$ the process is stationary and is termed a persistent  fGn. For $1<\beta<3$ the process is  non-stationary and termed an fBm. As a short-memory alternative we shall also consider the AR(1) process which has an exponentially decaying ACF and is completely characterized by the one-time-lag autocorrelation $\phi$ \citep{Storch}.

Significance of linear trends under various null models, some exhibiting LRM, was studied by \cite{CohnandLins} in the context of northern-hemisphere temperature data.  One of their main points was that trends classified as statistically significant under a short-memory null hypothesis might end up as insignificant under an LRM hypothesis. Here we will consider  the instrumental data record HadSST3 for global ocean temperature  \citep{kennedy2011} and the land temperature record HadCRUT3 \citep{jones2012}.  These records are sea-surface and land-air temperature anomalies relative to   the period 1961-90, with monthly resolution from 1850 to date. The analysis is made using a trend model which contains a linear plus a sinusoidal trend, although the methodology developed works for any parametrized trend model.  We test this model against the null model that the full temperature record is a realization of an AR(1) process, an fGn, or an fBm (the fBm model is of interest only for the strongly persistent ocean data). 

The significance tests are based on generation of an ensemble of synthetic realizations of the null models;  AR(1) processes ($\phi<1$), fGns ($0<\beta<1$), and fBms   ($1<\beta<3$). Each realization is fully characterized by a pair of parameters; $\theta\equiv (\sigma,\phi)$ for AR(1) and $\theta\equiv (\sigma,\beta)$ for fGn and fBm, where $\sigma$ is the standard deviation of the stationary AR(1) and fGn processes and the standard deviation of the differenced fBm. For an LRM null model the estimated value of $\hat{\beta}$   depends on which null model (fGn or fBm) one  adopts. As we will show below, for ocean data, it is not so clear whether an fGn or an fBm is the most proper model \citep{lennartz2009,rypdalJGR2013}, so we will test the significance of the trends under both hypotheses. 

Technically, we make use of the R package by \citet{mcleod2007}  to generate synthetic fGns and to perform a maximum-likelihood estimation of $\beta$. Synthetic fBms  with memory exponent $1<\beta<3$ are produced by generating an fGn with exponent $\beta-2$ and then forming the cumulative sum of that process. This is justified because the one-step differenced fBm with $1<\beta<3$ is an fGn with memory exponent $\beta-2$ \citep{beran1994}. Maximum-likelihood estimation of  $\beta$ for synthetic fBms and observed data records modeled as an fBm is done by forming the one-time-step increment (differentiation) process, estimate  the memory exponent $\beta_\text{incr}$  for that process and find $\beta=\beta_\text{incr}+2$. There are some problems with this method when $\beta \approx 1$. Suppose we have a data record (like the global ocean record) and we don't know whether  $\beta <1$ or $\beta >1$. For all estimation methods there are large errors and biases for short data records of fGns/fBms for $\beta\approx 1$ \citep{rypdalJGR2013}. This means that there is an ambiguity as to whether a record is a realization of an fGn or an fBm when we obtain estimates of $\beta$ in the vicinity of 1. For the MLE method this ambiguity becomes apparent from Figure 2. Here we have plotted the MLE estimate $\hat{\beta}$ with error bars for an ensemble of realizations of fGns (for $0<\beta<1$) and of fBms ($1<\beta <2$) with 2000 data points. The red symbols are obtained by adopting an fGn model when $\beta$ is estimated. Hence, for $\beta >1$ we find the estimate $\hat{\beta}$ from a realization of an fBm with a model that assumes that it is an fGn. It  would be expected that the analysis would give $\hat{\beta} \approx 1$ for  an fBm, but we observe that it gives $\hat{\beta}$ considerably less than 1 in the range $1<\beta<1.4$, so if we observe a $\hat{\beta}$ in the vicinity of 1 by this analysis we cannot know whether it is an fGn or an fBm. The ambiguity remains by estimating with a model that assumes that the record is an fBm, because this yields a corresponding positive bias as shown by the green symbols when the record is an fGn. This ambiguity seems difficult to resolve for  ocean data  as short as the monthly instrumental record.

\begin{figure}
\includegraphics[width=.45\textwidth]{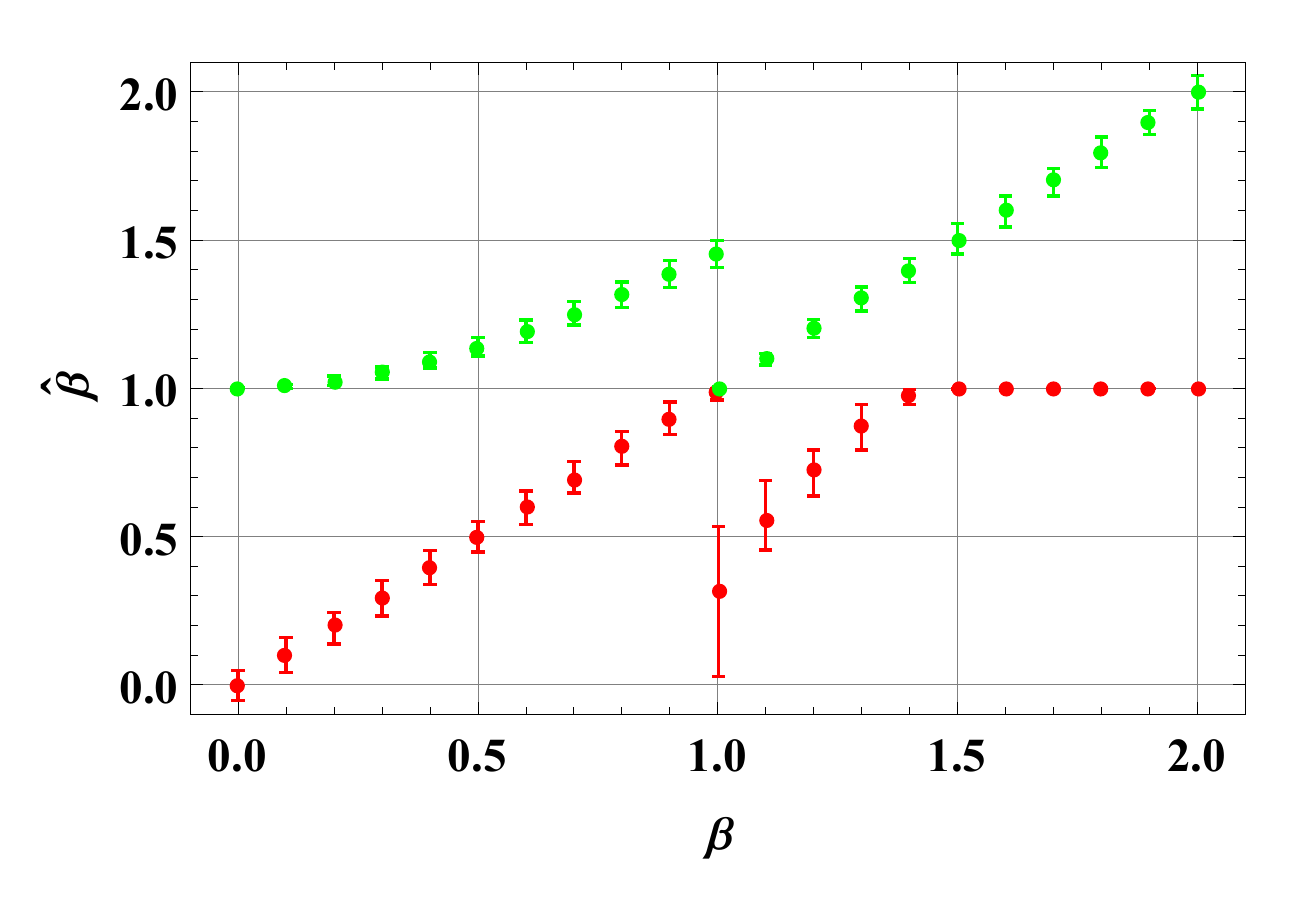}
\caption{The red symbols and 95\% confidence intervals represent the maximum-likelihood estimate $\hat{\beta}$ for realizations of   fGns/fBms with memory parameter $\beta$ by adopting an fGn model. Hence, for $\beta >1$ we find the estimate $\hat{\beta}$ from a realization of an fBm with a model that assumes that it is an fGn. The green  symbols represent the corresponding estimate by adopting an fBm model, i.e., for $\beta<1$ we we find the estimate $\hat{\beta}$ from a realization of an fGn with a model that assumes that it is an fBm. ``Adopting an fBm model''  means that the synthetic record is differentiated, then analyzed as an fGn by the methods of \cite{mcleod2007} to obtain $\hat{\beta}_{\text{incr}}$,  and then finally $\beta=\hat{\beta}_{\text{incr}}+2$.} 
\label{fig:errormle}
\end{figure}
 
The standard method  for establishing a trend in time-series data is to adopt a parametrized model $T(A;t)$ for the trend, e.g., a linear model $A_1+A_2t$ with parameters $A=(A_1,A_2)$, and estimate the model parameters  by a least-square fit of the model to the data. Another method, which brings along additional meaning to the trend concept, is the MLE method. This method adopts a model for the stochastic process; $x(t)=T(A;t)+\sigma w(t)$, where $w(t)$ is a correlated or uncorrelated random process and establishes the set of model  parameters $A$ for which the likelihood of the stochastic model to produce the observed data attains its maximum. The method applied to uncorrelated and Gaussian noise models is described in many standard statistics texts \citep{Storch}, and its application to fGns is described in \cite{mcleod2007}. If $w(t)$ is a Gaussian,  independent and identically distributed (i.i.d.) random process, the MLE is equivalent to the least square fit. If $w(t)$ is a strongly correlated (e.g., LRM) process, and the trend model provides a poor description of the large-scale structures in the data, MLE may assign more weight to the random process (greater  $\sigma$) than the least-square method. On the other hand, if the trend model is chosen such that it can be fitted to yield a good description of the large-scale structure, the parameters estimated by the two methods are quite similar, even if $w(t)$ used in the MLE method is an LRM process. In this case  we can use least-square fit to establish the trend parameters without worrying about whether the residual noise obtained after subtracting the estimated trend can be modeled as a Gaussian, i.i.d. random process. 

In the following, we make  some definitions and outline the methodology we adopt to assess the significance of the estimated trend. The method is based on standard hypothesis testing, where the trend hypothesis (termed the ``alternative hypothesis'') is accepted (although not verified, which is stronger) by rejection of a ``null hypothesis.'' Failure of rejection of the null hypothesis implies failure of acceptance of the alternative hypothesis, and hence the trend will be characterized as insignificant under this null hypothesis. Hence, it is clear that the outcome of the significance test will depend on the choice of alternative hypothesis (trend model) as well as on the null hypothesis (noise model).


Let us take the pragmatic point of view that a trend is a simple and slowly varying (compared with a predefined time scale $\tau$)  function $T(A;t)$ of $t$, parametrized by  the trend coefficients $A=(A_1,\ldots,A_n)$. It is also required that for the optimal choice of parameters, $A=\hat{A}_{\text{obs}}$ the trend $T(\hat{A}_{\text{obs}};t)$ makes a good fit to the large-scale structure of the data record. In practice, this means that the trend should be  close to a low-pass filtered version of the signal, for instance a moving average over time-scale $\tau$. The trend is significant with respect to a particular null model if the fitted $T(\hat{A}_{\text{obs}};t)$ is very unlikely to be realized in an ensemble of fits  $T(\hat{A};t)$ to realizations of the null model. 

\begin{description} 
\item {\em Remark 1}: There is an infinity of measures that one may use to reject the null model, given the data, which use no information about the trend model $T(A;t)$.  For instance, the structure of the record on  time  scales $<\tau$  could be inconsistent with the null hypothesis, while the trend could be consistent with it. In that case we would judge the trend as insignificant, although the null model is rejected by the observed data record. 
\end{description}

\begin{description}
\item {\bf The alternative hypothesis} can be formulated as follows:  The observed record $x(t)$ is a realization of the stochastic process
\begin{linenomath*}
\begin{equation}
T(A;t)+\sigma w(t), \label{altmodel}
\end{equation}
\end{linenomath*}
where the trend $T(A;t)$ is a specified function of  $t$ depending on the trend coefficients $A=(A_1,\ldots,A_n)$, and $w(t)$ is a Gaussian stationary random process of unit variance. These coefficients are estimated from a least-square fit to $x(t)$ and have the values $\hat{A}_{\text{obs}}$. We assume that the trend model can be fitted so  well to the data that MLE-estimates of $A$ with different noise models (white noise vs. strongly persistent fGn) give approximately the same $\hat{A}_{\text{obs}}$. It is part of the alternative hypothesis that the true value of  $A$ is close to the estimated  value $\hat{A}_{\text{obs}}$. 

\item {\em Remark 2}: Without specifying that $A\approx \hat{A}_{\text{obs}}$ the alternative hypothesis will not set a criterion that can be used to reject the  null hypothesis.

\item{\bf The null hypothesis} states that the record $x(t)$ is a realization of a stochastic process
\begin{linenomath*}
\begin{equation}
\varepsilon(\theta;t), \label{nullmodel}
\end{equation}
\end{linenomath*}
with certain properties to be specified (e.g., the process is AR(1), fGn, or fBm). Like for the alternative hypothesis, the parameters $\theta$ should be restricted to be close to the values $\hat{\theta}_{\text{obs}}$ found from estimating it from fitting the null model (\ref{nullmodel}) to the data record by means of MLE. How close will be discussed in Remark 4.

\item {\em Remark 3}: The properties of the null model for time scales $<\tau$ are irrelevant (see Remark 1), so a test of the null model should ignore these scales.

\item{\bf The Monte Carlo null ensemble} is the collection of realizations $x_i(\theta)\,, i=1,2,\ldots, $ of the null model process (\ref{nullmodel}).

\item {\em Remark 4}: The best choice of null model would be to utilize all our possible knowledge about the true parameter set $\theta$. This implies considering $\theta$ as a random variable, and hence  a Bayesian approach \citep{Gelman2004}. We generate the null ensemble by drawing $\theta$ from the conditional distribution $P(\theta|\hat{\theta}_{\text{obs}})$, i.e., the probability that the ``real" parameters of the observed process are $\theta$ given that the estimated parameters from the observed data are $ \hat{\theta}_{\text{obs}}$. One way of establishing this distribution is to generate an ensemble of realizations of the noise process with $\theta$ varied  in the relevant range $\theta\approx \hat{\theta}_{\text{obs}}$ and establish the conditional distribution $P(\hat{\theta}|\theta)$.   From Bayes' theorem one then has $P(\theta|\hat{\theta})=P(\hat{\theta}|\theta)P(\theta)/P(\hat{\theta})$. By setting $\hat{\theta}=\hat{\theta}_{\text{obs}}$, and assuming a flat prior distribution $P(\theta)$ in the relevant range in the vicinity of $\theta_{\text{obs}}$, we the find $P(\theta|\hat{\theta}_{\text{obs}})=P(\hat{\theta}_{\text{obs}}|\theta)$.

\item {\em Remark 5:} As an alternative to the Bayesian ideas described in remark 4 one could employ a frequentist approach. This means that we assume that the null model has a fixed true parameter value $\theta$. This parameter value is unknown, and the strategy is to create the Monte Carlo null ensemble $x_i(\hat{\theta}_{\text{obs}})\,, i=1,2,\ldots, $ using the $\theta$-values estimated from the observed data. We must then take the uncertainty in the $\theta$-estimates into account, since $\hat{\theta}_{\text{obs}}$ may deviate from the true $\theta$. This estimation error can be quantified using the bootstrap method, which assumes that the error in the parameter estimates in the null model with parameters $\theta$ can be well approximated by the corresponding errors for the null model with parameters $\hat{\theta}_{\text{obs}}$. When estimation errors are quantified one can easily adjust for these in the hypothesis tests.

\item{\bf Pseudotrend estimates $\hat{A}^{(i)}$} are the coefficients obtained by least-square fit of the trend model $T(A;t)$ to the realizations $x_i(\theta;t)$ of the null ensemble.

\item{\bf Pseudotrend distribution} is the $n$-dimensional PDF $P(\hat{A})$ over the null ensemble.

\item {\bf Null-hypothesis confidence region} is the region $\Omega$ in $n$-dimensional $A$-space for which $P(A)>P_{\text{thr}}$,  where $P_{\text{thr}}$ is chosen such that $\int_{\Omega}P(A)\, \mathrm{d}A=0.95$.

\item{\bf Significance of the trend model}  is established if the null hypothesis is rejected, e.g., the full $n$-dimensional trend is 95\% significant if $\hat{A}_{\text{obs}}\notin \Omega$.

\item {\em Remark 6}: If the null hypothesis is rejected by this procedure, we are rejecting only those aspects of the null model that are relevant to the full trend model, i.e., the trend model (alternative hypothesis) produces trend coefficients $\hat{A}_{\text{obs}}$ that give a good fit to the large-scale structure of the data, while it is very improbable that the null model can produce $\hat{A}$ in the vicinity of $\hat{A}_{\text{obs}}$. 

\end{description}
 We will apply the method  to global temperature record using the following trend model:
\begin{linenomath*}
\begin{equation}
T(A;t)=\delta +A_1t+A_2\sin (2\pi f t+\varphi ).
\label{eq:trend}
\end{equation}
\end{linenomath*}

\begin{table}
\caption{Estimated noise parameters $\hat{\theta}_{\text{obs}}$ from  the null hypotheses in  \eqref{nullmodel} and trend parameters $\hat{A}_{\text{obs}}$  estimated from the trend model  \eqref{eq:trend}.\tablenotemark{a}}
\centering
\begin{tabular}{l c c c c c c c c}
\hline
&AR(1)&\multicolumn{2}{ c }{fGn}&\multicolumn{2}{ c }{fBm}&\multicolumn{3}{c}{Trend}\\
&$\hat{\tau}_{\text{obs}}$&$\hat{\beta}_{\text{obs}}$&$\hat{\sigma}_{\text{obs}}$&$\hat{\beta}_{\text{obs}}$&$\hat{\sigma}_{\text{obs}}$&$\hat{A}_{1,\text{obs}}$&$\hat{A}_{2,\text{obs}}$&$\hat{T}_{\text{obs}}$\\
\hline
Ocean&21.3&0.994&0.25&1.45&0.086&4.21&0.128&69.7\\
Land&3.43&0.654&0.49&&&6.34&0.186&73.4\\
\hline
\end{tabular}
\label{tab:trend}
\tablenotetext{a}{The units for the trend estimation are months for $\hat{\tau}_{\text{obs}}$, $10^{-3}$~$^\circ$C/yr for $\hat{A}_{1,2,\text{obs}}$, and yr for the oscillation period $\hat{T}_{\text{obs}}$.}
\end{table}

This is a simplified version of the models used in several works by N. Scafetta (e.g.,  in \cite{Scafetta2011,Scafetta2012}) and the oscillation is supposed to model the 60-yr cycle observed in the instrumental record \citep{schlesinger1994}. The model contains five parameters to be estimated  from the observed temperature records by least-square fit. The frequency $f$ estimated for land and ocean records are slightly different, but both correspond to a period  close to 60 yr.  When estimating pseudotrends it has little meaning to let $f$ be a free parameter, since the synthetic noise records contain no preferred frequencies. We therefore fix $f$ equal to the value estimated from the observed records. Of the estimated pseudotrend coefficients $(\hat{A}_1,\hat{A}_2, \hat{\delta}, \hat{\varphi})$ only $(\hat{A}_1,\hat{A}_2)$  quantify the strength of the trend, so the relevant pseudotrend distribution to establish is $P(\hat{A}_1,\hat{A}_2)$ irrespective of the values of irrelevant parameters $(\hat{\delta},\hat{\varphi})$. 
Table 1 shows the estimated $\hat{\theta}_{\text{obs}}$ according to the null model in (\eqref{nullmodel}) using AR(1), fGn and fBm as the stochastic process $\varepsilon(\theta;t)$. Also in this table are the estimated trend parameters $(\hat{A_1},\hat{A_2},\hat{T})_{\text{obs}}$ from applying the trend model in  \eqref{eq:trend}, where $\hat{T}_{\text{obs}}=1/\hat{f}_{\text{obs}}$ is the estimated period of the oscillatory trend.

\begin{figure}
\noindent \includegraphics[width=.5\textwidth]{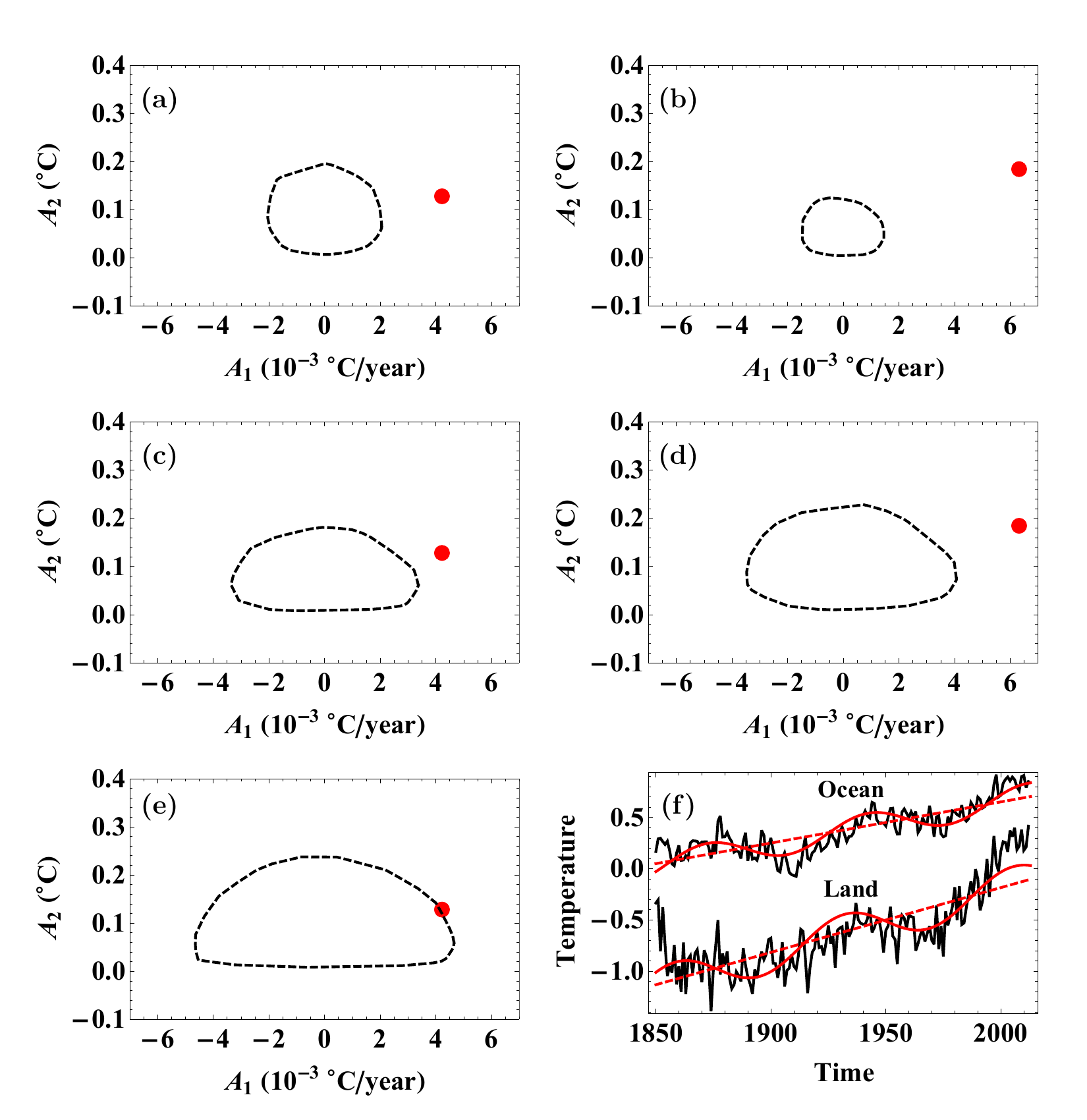}
\caption{In panels (a-e) the red dots represent the estimated trend coefficients $(\hat{A}_1,\hat{A}_2)_{\text{obs}}$ and the dashed, closed curve the 95\% confidence contour  of the distribution $P(\hat{A}_1,\hat{A}_2)$. (a): ocean data and AR(1) null model. (b): land data and AR(1) null model. (c):  ocean data and fGn null model. (d): land data  and fGn null model. (e):  ocean data and fBm null model. (f): Black curves: The global ocean and land temperature records. Red  curves: the linear and sinusoidal trends.} 
\label{fig:oceantrend}
\end{figure}

The results of the analysis are shown in Figure 3. We observe that the  trend parameters $(\hat{A}_1,\hat{A}_2)_{\text{obs}}$ are outside the null-hypothesis 95\% confidence region for all three noise models and for ocean as well as land records. But we also observe that the significance is more evident for land than for ocean, and is reduced as more strongly persistent noise models are employed. For the fBm model applied to ocean data the trend is barely outside the 95\% confidence region.

It is the full n-dimensional trend model that is accepted by this test, but something can also be said about the separate  significance of the individual trends represented by the individual trend coefficients from the pseudotrend distribution  $P(\hat{A}_1,\hat{A}_2)$. For the AR(1) and fGn null models it is apparent from Figure 3(a)-(d) that the linear trend is highly significant since $\hat{A}_{1,\text{obs}}$ is located far to the right of the confidence region. On the other hand, except for the AR(1) model applied to land data in Figure 3(b), $A_{2,\text{obs}}$ is not totally above the confidence region. This means that the linear pseudotrends observed in the null ensemble has negligible chance of getting near the observed trend, while there is some chance to find oscillatory trends in the null ensemble which are as large as $\hat{A}_{2,\text{obs}}$. The significance of those separate trends against these null models is determined by forming the separate one-dimensional PDFs, $P(\hat{A}_1)\equiv \int P(\hat{A}_1,\hat{A}_2)\mathrm{d}\hat{A}_2$ and $P(\hat{A}_2)\equiv \int P(\hat{A}_1,\hat{A}_2)\mathrm{d}\hat{A}_1$ and form the confidence intervals in the standard way. In Figure 4 we have formed the corresponding one-dimensional cumulative distribution functions (CDFs) from the two-dimensional PDFs for ocean data shown in Figure 3(a), (c), and (e). We observe that the linear trend is significant for the AR(1) and fGn null models, but barely significant for the fBm model. The oscillatory trend is insignificant for all models. 
The corresponding CDFs for land data are shown in Figure 5. The linear trend is even more significant than for ocean data,  while the oscillatory trend is significant for the AR(1) model, but barely significant for the fGn model.

One important lesson to learn from this  analysis is that the stronger persistence in the ocean temperature record  makes it harder to detect significant trends as compared to the land record. This is contrary to the common belief that the higher noise levels on short time scales in land records will make trend detection more difficult in these records. Another is that the land record analysis establishes beyond doubt that there is a significant global linear trend throughout the last century, and that the reality of an oscillatory trend is probable, but not beyond the 95\% confidence limit.

\begin{figure}
\noindent \includegraphics[width=.5\textwidth]{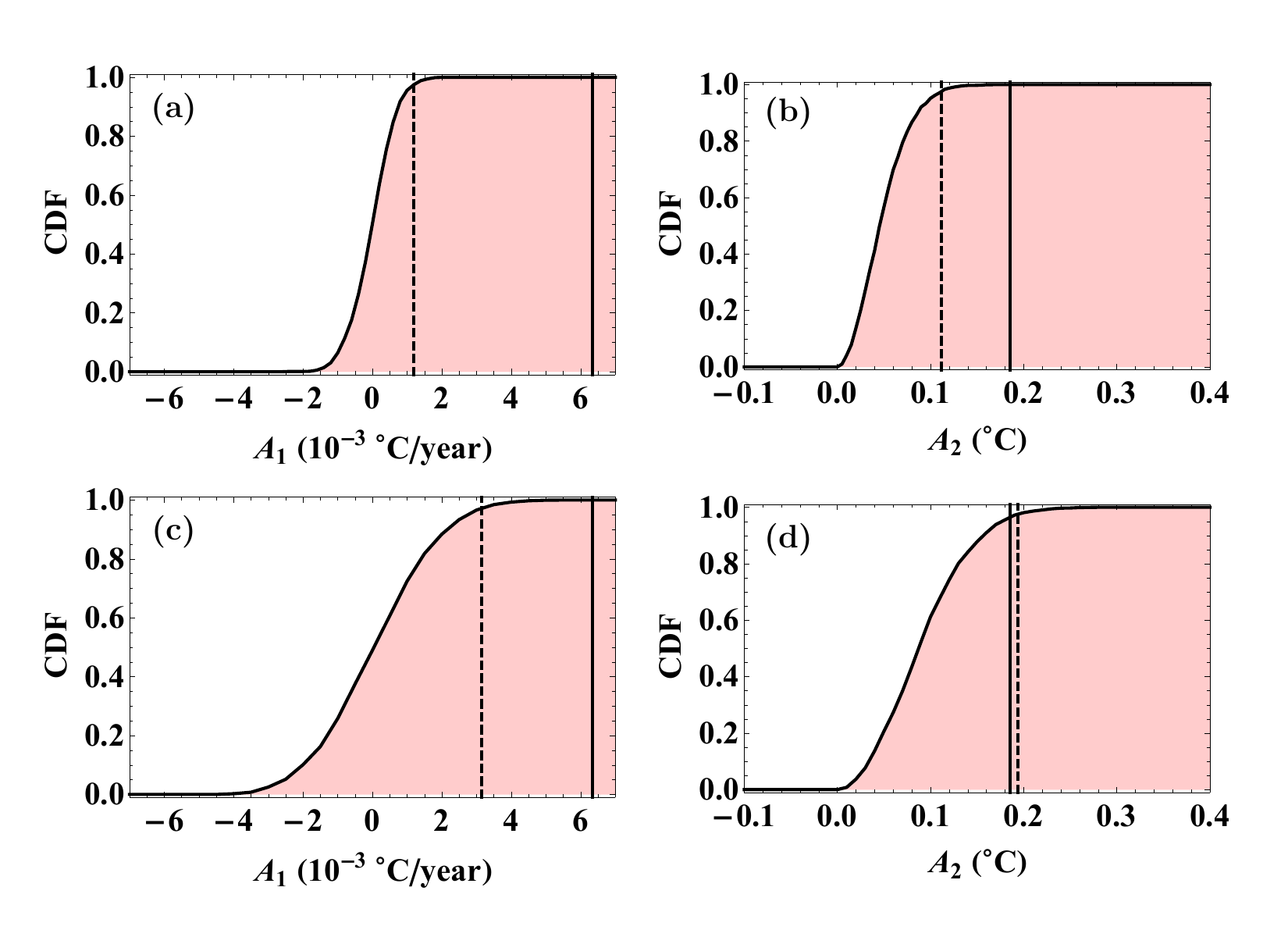}
\caption{Curved lines are CDFs for trend coeffecients $\hat{A}_1$ and $\hat{A}_2 $ established from the null model ensemble for land data.  Vertical dashed line marks the upper 95\% confidence limit. Vertical solid lines  mark $\hat{A}_{1,2,\text{obs}}$. (a) and (b): AR(1) null model. (c) and (d): fGn null model.}
\label{fig:cdfland}
\end{figure}

\section{Sharpening and Evaluating the Null Hypothesis}
In a Bayesian spirit, it would be appropriate to investigate the oscillatory trend further  by including the linear trend as an established fact and construct a sharper null model;

\begin{linenomath*}
\begin{equation}
\hat{\delta}_{\text{obs}} +\hat{A}_{1,\text{obs}}t+\varepsilon(\theta;t).
\label{newnullmodel}
\end{equation}
\end{linenomath*}

\begin{table}
\caption{Estimated noise parameters $\hat{\theta}_{\text{obs}}$ from  the new null hypotheses in \eqref{newnullmodel}. The units are same as in Table 1.}
\centering
\begin{tabular}{l c c c}
\hline
&AR(1)&\multicolumn{2}{ c }{fGn}\\
&$\hat{\tau}_{\text{obs}}$&$\hat{\beta}_{\text{obs}}$&$\hat{\sigma}_{\text{obs}}$\\
\hline
Land&2.04&0.584&0.391\\
\hline
\end{tabular}
\label{tab:newnull}
\end{table}

We now first  estimate a new $\hat{\theta}_{\text{obs}}$ by fitting the new null model (\ref{newnullmodel}) to the observed land record. The new  estimated noise parameters are shown in Table 2. Then we produce a new null ensemble of records from the null model by drawing $\theta$ from the conditional distribution $P(\theta|\hat{\theta}_{\text{obs}})$ as explained in Remark 4. Finally we fit the trend model (\ref{eq:trend}) to each realization in the ensemble and form $P(\hat{A}_1,\hat{A}_2)$. The result is shown for land data and $\varepsilon(\theta;t)$ modeled as an fGn in Figure 6(a). The inclusion of the linear trend in the null model will imply that we shall fit $\varepsilon(\theta;t)$ to the record $\tilde{x}(t)\equiv x(t)-(\hat{\delta}_{\text{obs}} +\hat{A}_{1,\text{obs}}t)$ rather than to $x(t)$. Since we already have established that $x(t)$  contains a significant linear trend the variability of $\tilde{x}(t)$ may be considerably less than the variability of $x(t)$ and hence the new estimated noise parameters $\hat{\theta}_{\text{obs}}$ may correspond to smaller $\hat{\sigma}_{\text{obs}}$ and $\hat{\beta}_{\text{obs}})$ than we obtained for the original null model. This reduction in noise parameters leads to  narrowing of $P(\hat{A}_1,\hat{A}_2)$, and a narrower CDF for the oscillation trend parameter $\hat{A_2}$,  as shown in Figure 6(b). The result is that this sharper test establishes that the oscillatory trend is also significant.

The long-range memory associated with fractional noises and motions allows larger fluctuations on long time scales that allows description of such variability as part of the noise background rather as trends. The implication is that variability which has to be described as  significant trends under  white-noise or  short-memory noise hypotheses are insignificant trends under an LRM null hypothesis. But how do we decide on the proper null hypothesis? One way to deal with this question is to use some estimator that characterizes the correlation structure of the observed record and compare that to the same estimator for different noise models. We have used the Mexican-hat wavelet variance  due to its conceptual simplicity and its ability to extract the noise component and eliminate obvious linear trends, and because it works without modification for both fGns and fBms \citep{rypdalJGR2013}. In Figure 7(a) and (b) we have plotted the wavelet variance versus time scale for the observed ocean and  land record and for ensembles of synthetic realizations of  short-memory AR(1) processes, fGns and  (for  ocean) fBms, with parameters estimated from the observed records by the MLE method. The first thing to notice is that the wavelet variance curve  for the ocean seems to fit reasonably well within the confidence limits for the fBm null model, but not that well for the fGn model and the AR(1) model. For the land record the wavelet-variance curve is also way outside the limits for both AR(1) and fGn. The reason for this is that the MLE estimates of $\hat{\theta}_{\text{obs}}$ attempts to fit the noise process to a record that consists of  a linear trend in addition to a noise, and hence overestimates the noise parameters. The wavelet variance will therefore be estimated from synthetic realizations of too strong and too persistent noise processes. The wavelet variance of the observed signal will be closer to that of the true noise component, because the method automatically removes the effect of  a linear trend. This is why the wavelet-variance curve of the observed signal is  below the confidence region for the null ensemble in Figure 7(b), and is an additional confirmation that the linear trend is real.

In Figure 7(c) and (d) we have plotted the wavelet variance curves of the linearly detrended observed records. They are identical to the wavelet variance for the full records, which demonstrates that the wavelet variance is insensitive to a linear trend. The  colored curves and confidence regions are produced from the null ensembles produced  from the new null model (\ref{newnullmodel}). As discussed above, this null model has weaker and less persistent noise, and this reduces the wavelet variances and bring the curves more in line with those obtained from the observed records. Now it appears that the observed wavelet variance is within the confidence limits for the LRM processes fGn/fBm, but somewhat outside those limits for the short-range AR(1) process, suggesting a preference for the LRM null models. These  features appear considerably clearer when residuals between the observed records and the deterministic records from the forced response models are analyzed by wavelet variance \citep{RR2013}, the reason being primarily that the small-scale features of the forced response in that case is subtracted, and hence this residual will give a better representation of  the internal climate-noise component. 

The differences between Figure 7(a), (b) and Figure 7(c), (d) show that the residual after linear detrending is much better described as a noise process than the undetrended record, and hence gives a clear indication that the new null model (\ref{newnullmodel}) is  better than the original model (\ref{nullmodel}).

\section{Conclusions}
In this paper we have attempted to classify the various possible ways to understand the notion of a trend in the climate context, and then we have focused on the detection of a combination of a rising and oscillatory trend in global ocean and land instrumental data when no information about the climate forcing is used. It is well known that the statistical  significance of the trends depends on the degree of autocorrelation (memory) assumed for the random noise component of the climate record \citep{CohnandLins,rybski2006,rybski2009}. It is also known that the linear trends are easier to detect and appear to be more significant in global than in local data \citep{lennartz2009}, although local records exhibit weaker long-term persistence than global records. Despite this fact, much effort is spent on establishing trends and their significance in data from local stations (e.g., \citep{Franzke2012}) with variable results. The failure of detecting consistent trends in local data records  reflects the tendency of  internal spatiotemporal variability to mask the trend that signals global warming, and  we believe therefore that investigation of such trends should be performed on globally averaged data. For global data records our study demonstrates very clearly that the long-range memory observed in sea-surface temperature data leads to lower significance  of detected trends compared to land data. This does not mean, of course, that the global warming signal and internal oscillations are not present in the local records or in the global ocean record. It is just not possible to establish the statistical significance of these trends from these records alone, since the large short-range weather noise in local temperatures and the slower fluctuations in ocean temperature  both reduce the possibilities of trend detection. Hence, one  needs to search for the optimal climate record to analyze for detection of  the global warming signal, and our results suggest that the global land temperature signal may be the best candidate for such trend studies. 

\begin{figure}
\noindent 
\includegraphics[width=.5\textwidth]{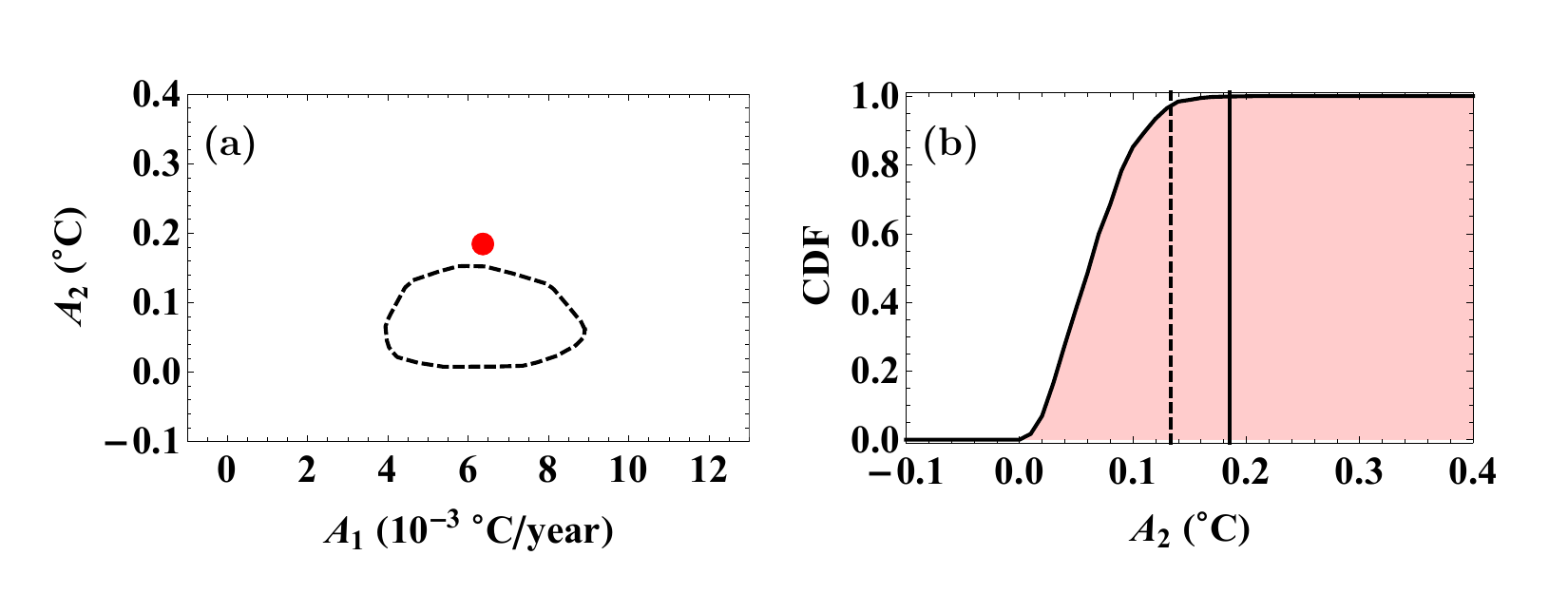}
\caption{(a): The 95\% confidence contour of the distribution $P(\hat{A}_1,\hat{A}_2)$ for land data obtained by the new null model (\ref{newnullmodel}) with $\varepsilon(\theta;t)$ an fGn process. (b): The CDF derived from  $P(\hat{A}_2)$ for this null model, with upper 95\% confidence limit marked as dotted vertical line.}
\label{fig:cdfnewnull}
\end{figure}

\begin{figure}
\noindent\includegraphics[width=.5\textwidth]{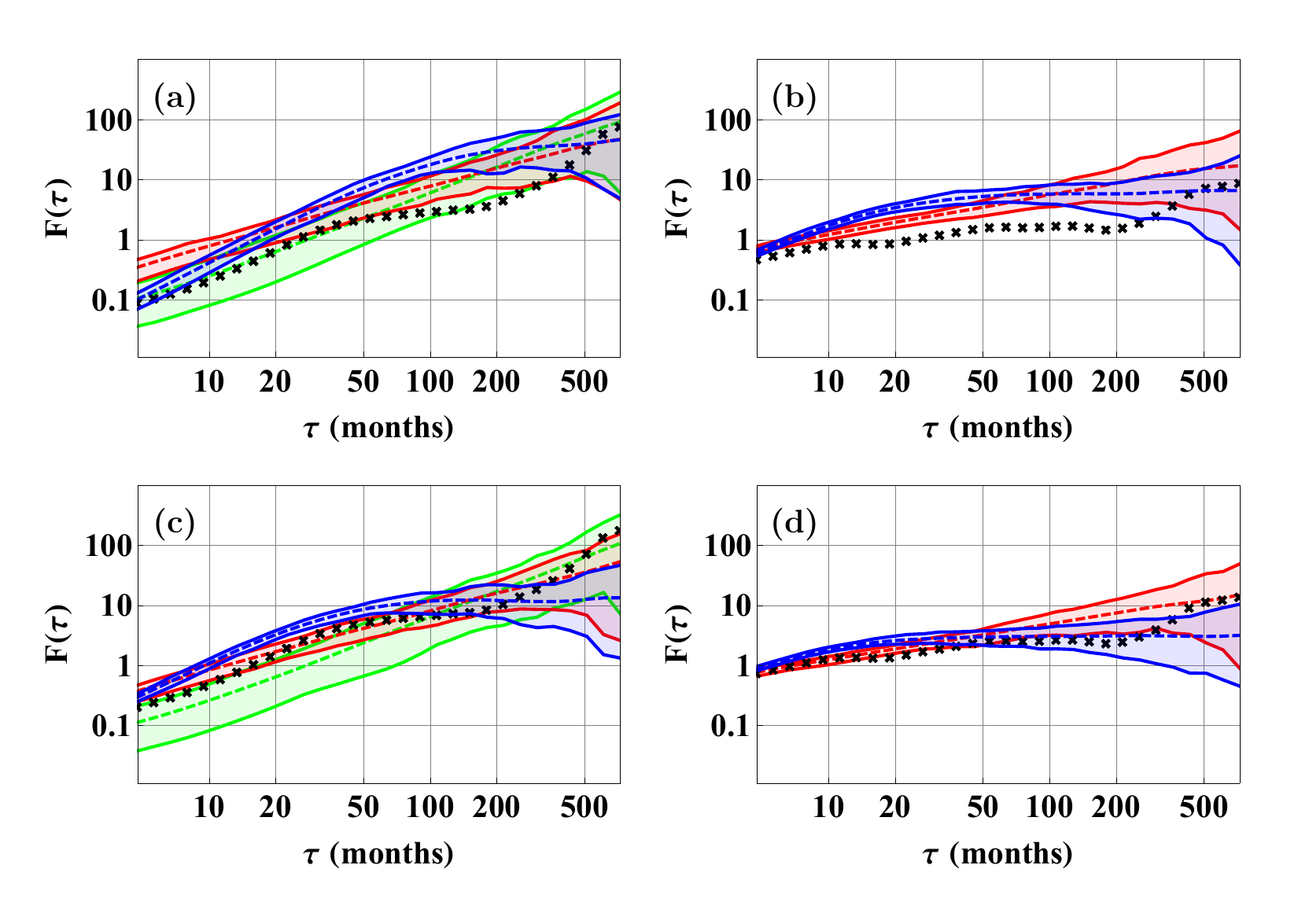}
\caption{Panels (a) and (b) show the wavelet variance versus time scale for the observed ocean (a) and land (b) records (black crosses) and  for ensembles of synthetic realizations of   AR(1) processes (blue), fGns (red) and  (for  ocean) fBms (green), with parameters estimated from the observed records by the MLE method. The shaded areas are 95\% confidence regions for these estimates. Panels (c) and (d) show the wavelet-variance curves of the linearly detrended, observed records (black crosses) and for the synthetic realizations of the processes generated from the new null model (\ref{newnullmodel}). }
\label{fig:waveletvariogram}
\end{figure}

While a linear trend is only marginally significant under the long-range memory null hypothesis in ocean data, it is clearly significant in land data. Hence, there should be no doubt about the significance of a global warming signal over the last 160 years even under null hypotheses presuming strong long-range persistence of the climate noise. 

There is some more doubt about the significance of a 60 year oscillatory mode in the global signal, as  shown in Figure 4(d) and 4(f) and Figure 5(d). By mean of a Bayesian iteration, however, utilizing the established significance of a linear trend to formulate a sharper null hypothesis, we are able to establish statistical significance of the oscillatory trend in the land data record. We believe this is an important result, because it means that we cannot dismiss this oscillation as a spontaneous random fluctuation in the climate noise background. By the analysis presented here  we cannot decide whether this oscillation is an internal mode in the climate system  or an oscillation forced by some external influence. Such insights can  be obtained from a generalization of the response model of \cite{RR2013} by employing information about the climate forcing, and will be the subject of a forthcoming paper. There are various published hypotheses about the nature of this oscillation. The least controversial is that this is a global manifestation of the Atlantic Multidecadal Oscillation (AMO) which is essentially an internal climate mode \cite{schlesinger1994}. Some authors go further and suggest that this oscillation is synchronized and phase locked with some astronomical influence \citep{Scafetta2011,Scafetta2012}. Although some of these suggestions seem very speculative, there are some quite well-documented connections between periodic tidal effects on the Sun from the motion of the giant planets and radioisotope paleorecord proxies for solar activity  on century and millennium time scales \cite{Abreu}. So far there exists no solid evidence that these, and multidecadal,  variations in solar activity have a strong influence on terrestrial climate, but the issue will probably be in the frontline of research on natural climate variability in the time to come.

 \appendix 





\begin{acknowledgments}
The authors are grateful to Ola L{\o}vsletten for illuminating discussions and comments.
\end{acknowledgments}

%
%

\end{article}

\end{document}